\documentclass[prd, twocolumn, nofootinbib, floatfix]{revtex4-1}

\usepackage[mathscr]{euscript}
\usepackage{amsmath}
\usepackage{graphicx}
\usepackage{dcolumn}
\usepackage{bm}
\usepackage{epsfig}
\usepackage{amssymb,latexsym,mathrsfs}
\usepackage{graphicx}
\usepackage{color}
\usepackage{hyperref}
\usepackage{float}
\usepackage{diagbox}
  
\definecolor{vividviolet}{rgb}{0.62, 0.0, 1.0}
\definecolor{amaranth}{rgb}{0.9, 0.17, 0.31}
\definecolor{palatinateblue}{rgb}{0.15, 0.23, 0.89}
\definecolor{brightpink}{rgb}{1.0, 0.0, 0.5}
\definecolor{cornflowerblue}{rgb}{0.39, 0.58, 0.93}
\definecolor{deepcarminepink}{rgb}{0.94, 0.19, 0.22}
\definecolor{radicalred}{rgb}{1.0, 0.21, 0.37}

\hypersetup{ linktoc=all,
    colorlinks, linkcolor={palatinateblue},
    citecolor={brightpink}, urlcolor={amaranth}
}

\newcommand{\be}{\begin{equation}}
\newcommand{\ee}{\end{equation}}
\newcommand{\bs}{\begin{split}} 
\newcommand{\bea}{\begin{eqnarray}}
\newcommand{\eea}{\end{eqnarray}}

\renewcommand{\d}[1]{\ensuremath{\operatorname{d}\!{#1}}}

\begin{document}

\title{The mirror at the edge of the universe: Reflections on an accelerated boundary correspondence with de Sitter cosmology}
\author{Michael R.R. Good${}^{1,2}$}
\email{michael.good@nu.edu.kz}
\author{Abay Zhakenuly${}^{1}$}
\email{abay.zhakenuly@nu.edu.kz}
\author{Eric V. Linder${}^{2,3}$}
\email{evlinder@lbl.gov}
\affiliation{${}^1$Physics Department, Nazarbayev University,
\\
Nur-Sultan, Kazakhstan\\
${}^2$Energetic Cosmos Laboratory, Nazarbayev University,\\ Nur-Sultan, Kazakhstan\\
${}^3$Berkeley Center for Cosmological Physics \& Berkeley Lab, University of California, Berkeley, CA, USA 
}

\begin{abstract} 
An accelerated boundary correspondence (ABC) is solved for the de Sitter moving mirror cosmology.  The beta Bogoliubov coefficients reveal the particle spectrum is a Planck distribution with temperature inversely proportional to horizon radius.  The quantum stress-tensor indicates a constant emission of energy flux consistent with eternal equilibrium, while the total energy carried by the particles remains finite. The curved spacetime transformation to flat spacetime with an accelerated boundary is illustrated, and also shown for Anti-de Sitter (AdS) spacetime. 
\end{abstract} 

\date{\today} 

\maketitle 

\section{Introduction} 

Understanding the thermodynamics and quantum particle production of de Sitter space \cite{GibbonsHawking,spradlin2001les,Balasubramanian_2001} is well-motivated both 
mathematically, since it is the maximally symmetric solution of Einstein's equations with a positive cosmological constant, and 
physically, since during the early universe, $t\lesssim 10^{-32}\,$s, inflation is approximately de Sitter \cite{Linde:2007fr} and 
with current cosmic acceleration the cosmos may be headed 
for a future de Sitter state. 

The static coordinate metric of de Sitter is given by 
\be ds^2 = -\left(1-\frac{r^2}{L^2}\right)dt^2 + \left(1-\frac{r^2}{L^2}\right)^{-1}dr^2+r^2d\Omega\,,\ee
with $d\Omega \equiv d\theta^2 + \sin^2\theta d\phi^2$.  
We  transform de Sitter space, with its horizon at $r=L$, to 
the analogous moving mirror model \cite{Davies:1976hi,Davies:1977yv} trajectory, with the accelerating 
boundary playing the role of the horizon, to study quantum particle 
production, i.e.\ the dynamical Casimir effect (see e.g.\ recent experimental proposals \cite{Chen:2020sir, Chen:2015bcg}). 

In Sec.~\ref{sec:dS} we derive the relation between the de 
Sitter metric, null sphere expansion, and the matching condition for accelerated motion.  
Section~\ref{sec:particles} computes the quantum particle spectrum and compares it to the late-time Schwarzschild mirror solution (the eternal black hole spectrum of the Carlitz-Willey mirror). 
We extend the mapping to Anti-de Sitter (AdS) space in Sec.~\ref{sec:AdS}.

\section{From de Sitter Metric to Acceleration} \label{sec:dS} 

A spherically symmetric, static metric 
\be \d s^2 = -f(r)\d t^2 + f(r)^{-1}\d r^2 + r^2\d\Omega\,, \ee 
corresponds to the inside static metric of the de Sitter 
expansion system for $f(r) \equiv f_L$ with 
\be f_L = 1-\frac{r^2}{L^2}\,.\label{metric} \ee 
While it looks similar to a black hole spacetime, here the observer lives in the inside, $r<L$, with a cosmological horizon at $r=L$.  The temperature seen by an inertial observer in de Sitter spacetime is 
\be T = \frac{1}{2\pi L}\,,\ee 
(see Appendix~\ref{appx} for a derivation). We set 
$G = \hbar = c  = 1$. 

For a double null coordinate system $(u,v)$, with $u = t-r^*$ and $v = t+r^*$, where the appropriate tortoise coordinate \cite{spradlin2001les} is
\be r^* = \int f^{-1}_L \,dr = \frac{L}{2}\ln \left|\frac{L+r}{L-r}\right| = L \tanh ^{-1}\frac{r}{L}\, ,\ee 
one has the metric for the geometry describing the inside region  $r<L$, 
\be \d s^2 =-f_L \; \d u \d v + r^2 \d\Omega\,.\ee 

The matching condition (see e.g.\ \cite{wilczek1993quantum,Fabbri}) with the flat exterior geometry, described by the exterior coordinates $U=T-r$ and $V=T+r$, is the trajectory of $r=0$, expressed in terms of the interior function $u(U)$ with exterior coordinate $U$: 
\be u(U) = v_0-2 L \tanh ^{-1}\frac{v_0-U}{2 L}\,.\ee 
This matching, $r^*(r=(v_0 -U)/2) = (v_0-u)/2$, happens along a light sphere, $v_0$.  Here $v_0 \pm 2 L \equiv v_H$ because $u\rightarrow \pm \infty$ at $U=v_H$. Without loss of generality we can set $v_0=0$.  The two horizons are at $v_H= \pm 2L$.  
The matching condition becomes 
\be u(U) = 2L \tanh^{-1} \frac{U}{2L}\,.\label{match}\ee 

The quantum field must be zero at $r=0$, ensuring regularity of the modes, such that the origin acts like a moving mirror in the $(U,V)$ coordinates.  Since there is no field behind $r<0$, the form of field modes can be determined, such that a $U\leftrightarrow v$ identification is made for the Doppler-shifted right movers.  We are now ready to analyze the analog mirror trajectory, $f(v) \leftrightarrow u(U)$, a known function of advanced time $v$. 

In the standard moving mirror formalism \cite{Birrell:1982ix} we study the massless scalar field in $(1+1)$-dimensional Minkowski spacetime (following e.g.\  \cite{Good_2015BirthCry}). From Eq.~(\ref{match}) the de Sitter analog moving mirror trajectory is 
\be f(v) = \frac{2}{\kappa}\tanh^{-1} \frac{\kappa v}{2}\,,\label{f(v)}\ee 
which is now a perfectly reflecting boundary in flat spacetime rather than the origin as a function of coordinates in curved de Sitter spacetime.  
Introduction of $\kappa \equiv L^{-1}$ is done to signal that we are now working in the moving mirror model with a background of flat spacetime, where $\kappa$ is the acceleration parameter of the trajectory. 

\begin{figure}[ht]
\centering 
\includegraphics[width=3.4in]{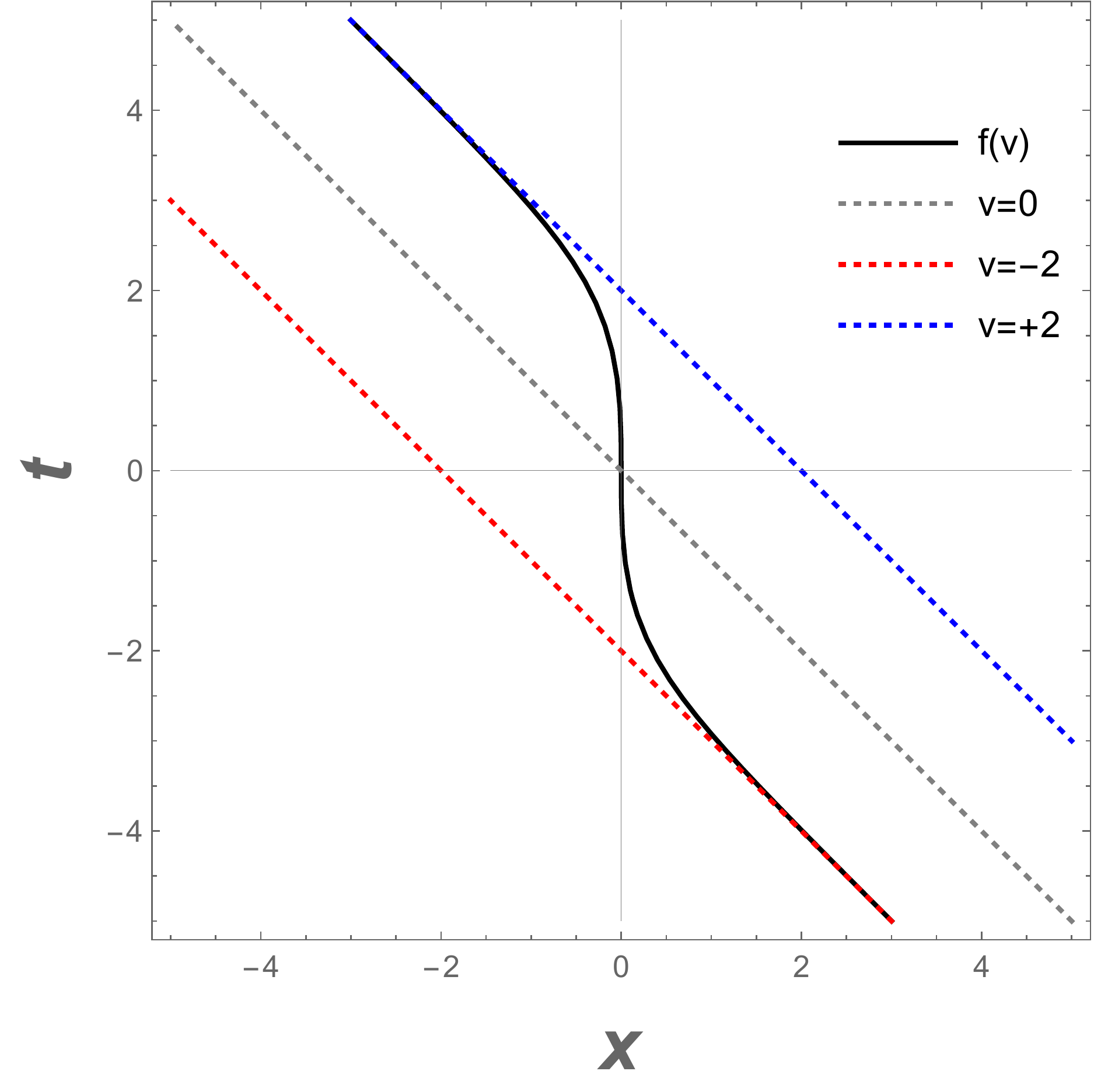} 
\caption{Trajectories, Eq.~(\ref{f(v)}), in a spacetime plot. The mirror is the thick black line. The horizons are at $\kappa v_H = \pm 2$ (blue and red  dotted lines) and the $v=0$ line is shown by the gray dotted line, where retarded time is $u=t-x$, and $v=t+x$ is advanced time. 
}
\label{fig:SpacetimePlot} 
\end{figure}

\begin{figure}[ht]
\centering 
\includegraphics[width=3.4in]{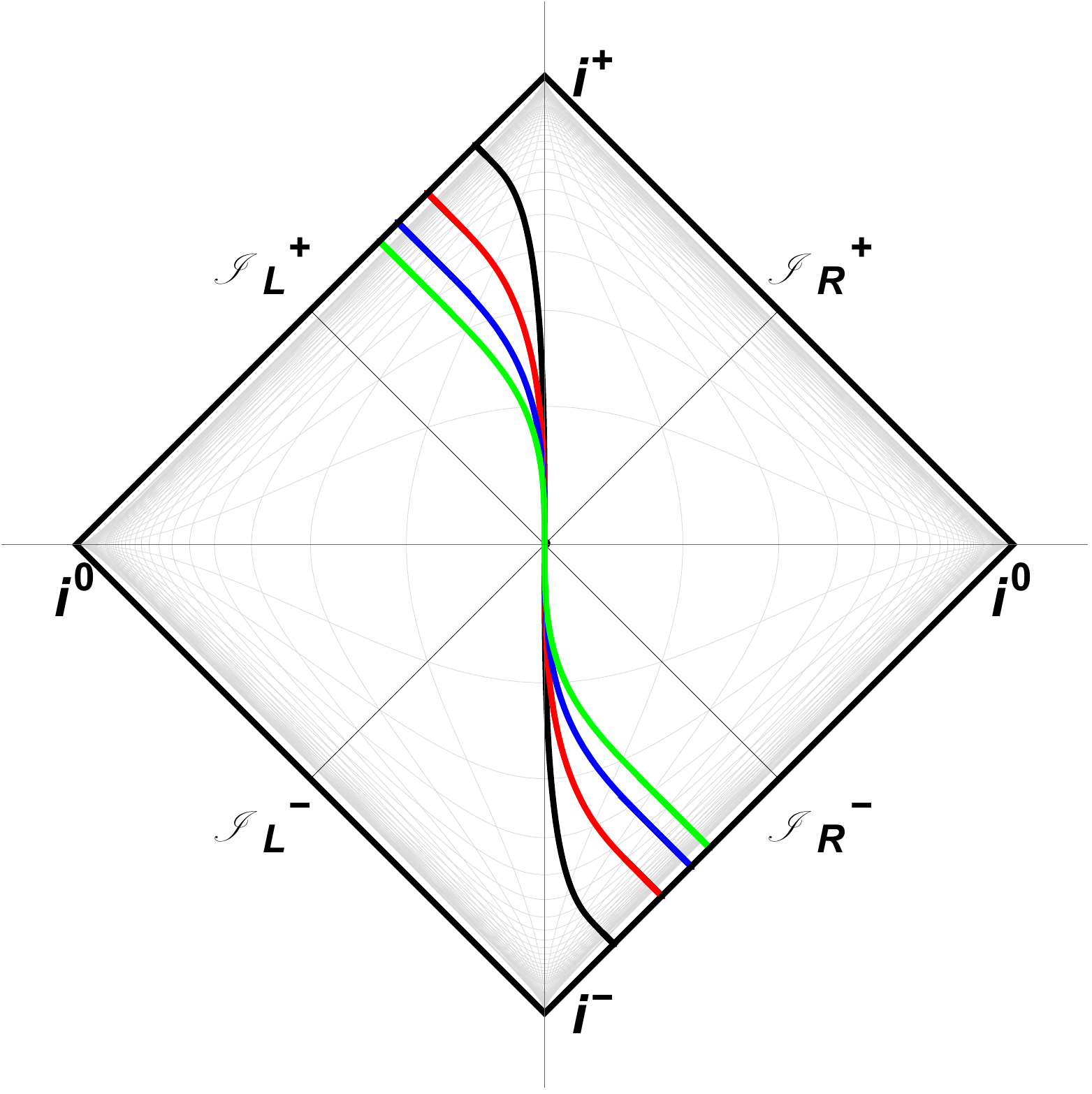} 
\caption{The class of trajectories Eq.~(\ref{f(v)}) in a Penrose conformal diagram. The colors correspond to different $\kappa = 1,2,3,4$: black, red, blue, green, respectively. The trajectories extend all the way out to the null future and past surfaces.  
}
\label{fig:PenrosePlot} 
\end{figure}

The horizons are $v_H=\pm 2/\kappa$, and so $\kappa v$ spans $-2 < \kappa v < +2$. 
The rapidity, in advanced time, 
$-2\eta(v) = \ln f'(v)$, is 
\be \eta(v) = \frac{1}{2} \ln \left(1-\frac{\kappa^2 v^2}{4}\right)\,.\label{rapidity}\ee
The rapidity asymptotes at $\kappa v = \pm 2$, i.e.\ the mirror approaches the speed of light at the horizons, $u\rightarrow \pm \infty$. The trajectory in spacetime coordinates is plotted as a spacetime plot in Fig.~\ref{fig:SpacetimePlot}.  A conformal diagram of the accelerated boundary is given in Fig.~\ref{fig:PenrosePlot}.  

The proper acceleration $\alpha = e^{\eta(v)}\eta'(v)$ is
\be \alpha(v)= -\frac{1}{2}\frac{\kappa^2 v}{ \sqrt{4-\kappa^2 v^2}}\,, \label{a(v)}\ee 
where prime denotes derivative with respect to the argument. 
Note the acceleration is zero at $v=0$, but approaches $\pm\infty$ 
near the horizons. The acceleration also takes on a simple form in 
terms of proper time, as discussed in Appendix~\ref{propertime}. 

The double divergence in the advanced time acceleration Eq.~(\ref{a(v)}) is arguably the main trait characterizing the de Sitter trajectory.  As a result, the de Sitter mirror possesses a double asymptotic null horizon in contrast to the single horizons of the Schwarzschild mirror \cite{Good_2017Reflections,Good_2017Reflections,Anderson_2017,Good_2017BHII} and the recently calculated trajectory of the extreme Reissner-Nordstr\"om mirror \cite{good2020extreme}.  Even the Reissner-Nordstr\"om mirror \cite{good2020particle}, (whose black hole has two horizons) has only one outer horizon relevant for its time-dependent particle production calculation. 
We will find the double horizons play an important role in the particle spectrum.

\section{Flux, Spectrum, and Particles} \label{sec:particles} 

For the de Sitter mirror, we find that the energy flux is constant, eternally. The radiated energy flux as computed from the quantum stress tensor is the Schwarzian derivative of Eq.~(\ref{f(v)}), \cite{Good_2017Horizonless}  
\be F(v)= \frac{1}{24\pi}\{f(v),v\}f'(v)^{-2},\ee
where the Schwarzian brackets are defined as
\be \{f(v),v\}\equiv \frac{f'''}{f'} - \frac{3}{2}\left(\frac{f''}{f'}\right)^2\,,\ee 
which yields 
\be F = \frac{\kappa^2}{48\pi}\,. \label{constantflux}\ee 
This result is indicative of thermal equilibrium and we next present the derivation of the accompanying Planck distribution.

The particle spectrum is given by the beta Bogoliubov coefficient, which can be found via \cite{Good_2017Horizonless} 
\be \beta_{\omega\omega'} = -\frac{1}{4\pi\sqrt{\omega\omega'}}\int_{v_H^-}^{v_H^+} \d v ~e^{-i \omega' v -i \omega f(v)}\left(\omega f'(v)-\omega'\right)\,,\label{betaint}\ee 
where $\omega$ and $\omega'$ are the frequencies of the outgoing 
and ingoing modes respectively. 
The result of the integration is 
\be \beta_{\omega\omega'}= \frac{2\sqrt{\omega \omega'}}{\kappa^2\sinh{\pi \omega/\kappa}} e^{-2 i \omega'/\kappa}M\left(1+i\omega/\kappa;2;4 i \omega'/\kappa\right), \label{eq:dSbeta} \ee  
where $M := \,_1F_1$ the confluent hypergeometric function, i.e.\ the Kummer function of the first kind. The same betas \cite{foo2020generating, Su2016} have been derived in the context of spacetime diamonds \cite{ Martinetti_2003}.

\begin{figure}[ht]
\centering 
\includegraphics[width=3.4in]{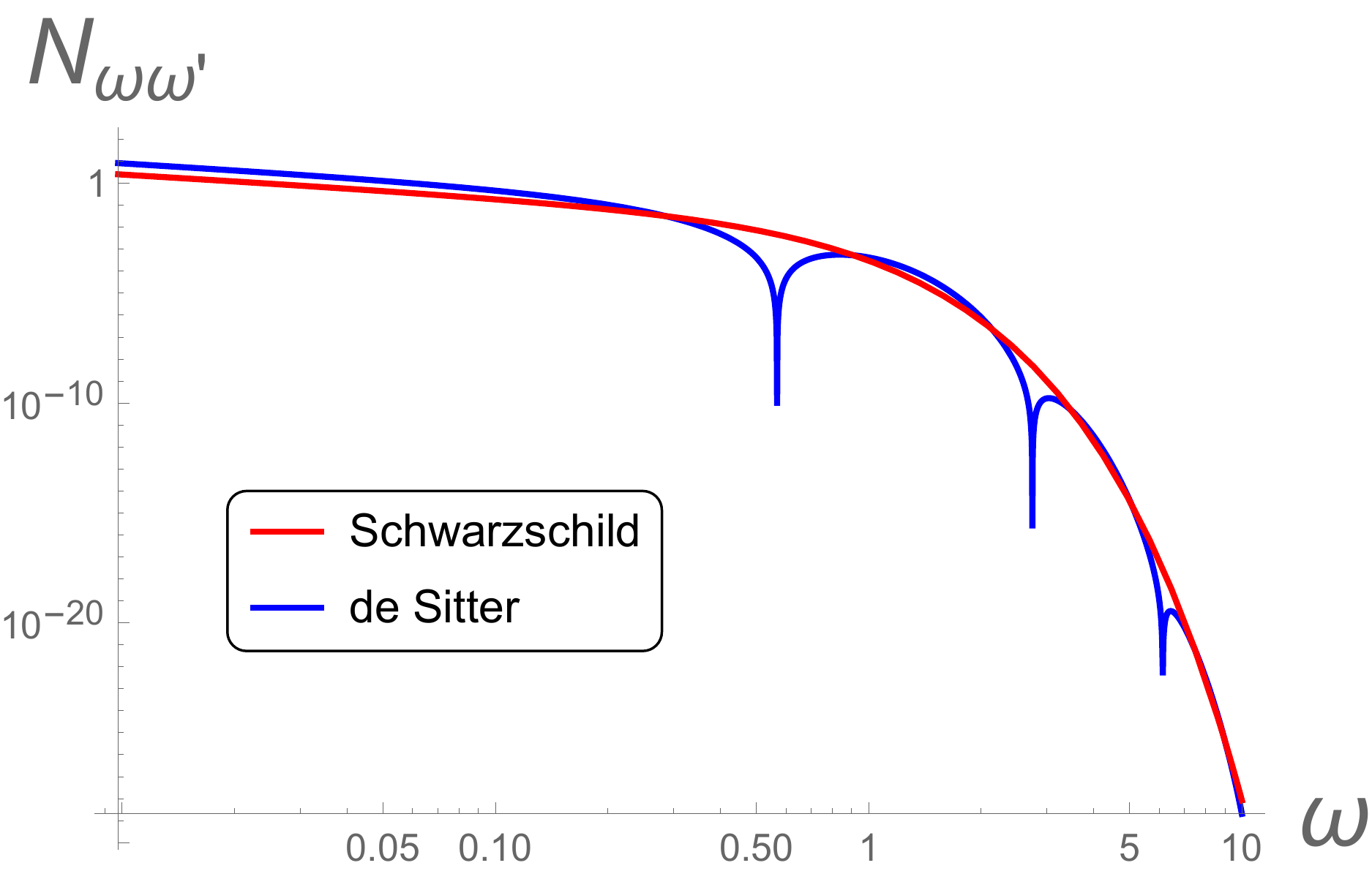} 
\caption{The mode spectra $N_{\omega\omega'} \equiv |\beta_{\omega\omega'}|^2$, setting $\omega'=1$ for illustration. The blue curve is the de Sitter case 
from Eqs.~(\ref{eq:dSbeta}-\ref{dSthermal}), while the red curve is the late time Schwarzschild result (or equivalent Carlitz-Willey 
mirror \cite{CW}) of Eq.~(\ref{CWthermal}). 
A larger pre-factor on the beta elevates the de Sitter spectra, offsetting the  
dips in frequency that are zeros in the mode spectrum due to Kummer's function, indicative of complete spectral absorption lines in the measure $|\beta_{\omega\omega'}|^2$. The zeros occur at $\omega = 0.57$, $2.74$, $6.14$, ad inf.  The destructive interference, like in the double-slit experiment, could be between the double-horizons, as field modes do not propagate freely asymptotically, i.e.\ $e^{i\omega v}$ has a boundary at $v=\pm v_H$. 
}\label{fig3} 
\end{figure}

\begin{figure}[ht]
\centering 
\includegraphics[width=3.4in]{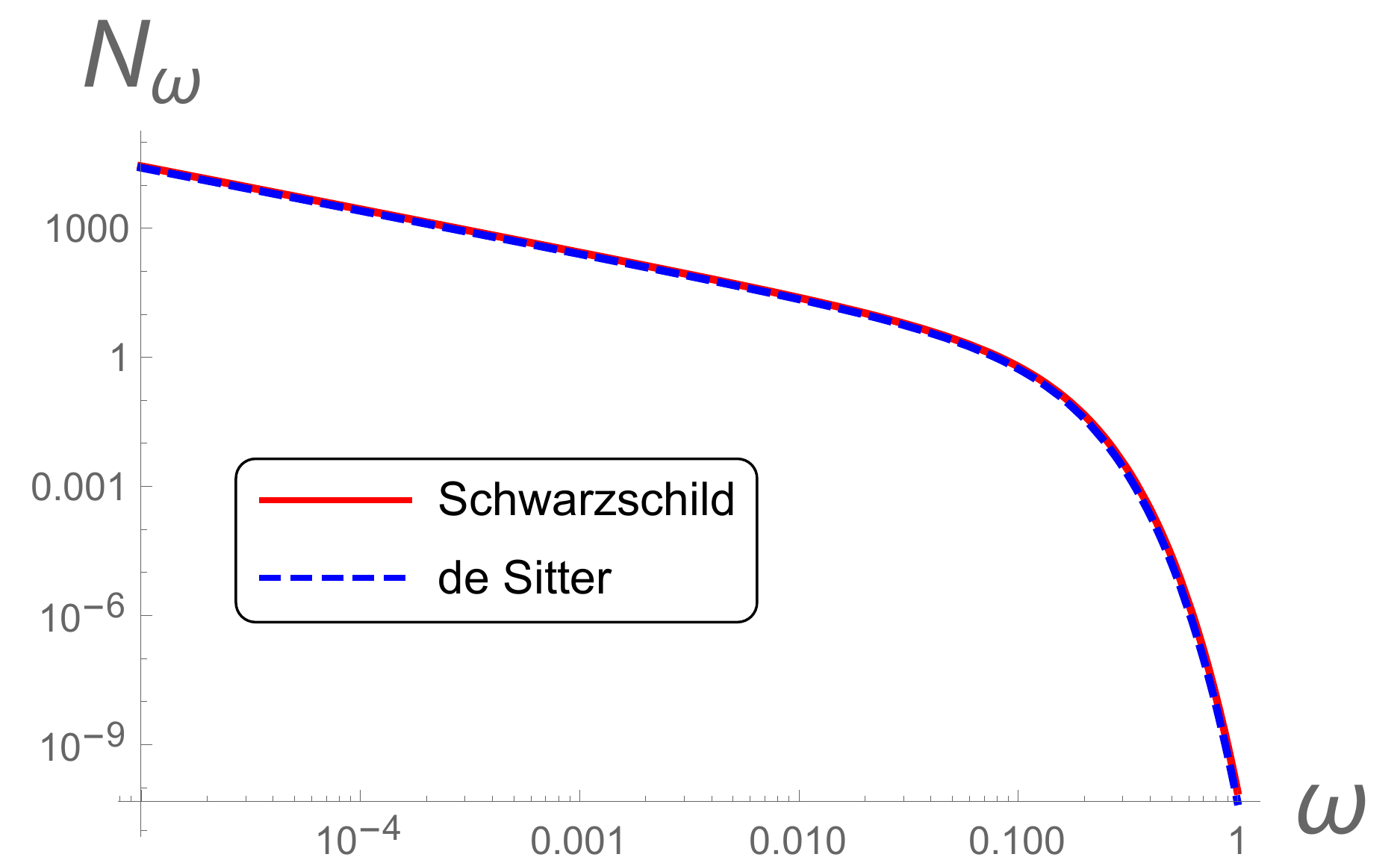} 
\caption{The particle count spectra $N_{\omega} =\int N_{\omega\omega'}d\omega'$. Blue and red curves are the same, showing that both de Sitter and late-time Schwarzschild solutions have thermal spectra at temperature $\kappa/2\pi$.  Here we have chosen $\kappa = 1/4$ for illustration.   
}\label{fig4} 
\end{figure}

To obtain the particle spectrum, we complex conjugate,
\be N_{\omega \omega'} \equiv |\beta_{\omega\omega'}^{\textrm{dS}}|^2\,, \label{dSthermal}\ee
giving the particle count per mode squared, plotted in Fig.~\ref{fig3}.  The spectrum $N_\omega$ is then 
\be N_\omega = \int_0^\infty  N_{\omega \omega'} d\omega'\,, \label{N(w)}\ee
plotted in Fig.~\ref{fig4}, illustrating graphically a thermal Planck particle number spectrum. Multiplying by the energy and phase space factors gives the usual Planck blackbody energy spectrum. 

Thermal behavior can also be seen analytically by the expectation value of the particle number, $\mathcal{N}_\omega$, via continuum normalization modes, 
\be \mathcal{N}_\omega \equiv \int_0^{+\infty} d \omega' \beta^{}_{\omega \omega'}\beta^{*}_{\omega_2 \omega'} = \frac{\delta(\omega-\omega_2)}{e^{2\pi \omega/\kappa}-1}\,.\ee  
We have used the textbook notation of \cite{Fabbri}, where the late-times Hawking case is done. As shown there, the delta function divergence (see also e.g. \cite{Minoru1979}) can be removed easily by using finite normalization  wave packet modes  \cite{Hawking1975}. The de Sitter calculation is not as straightforward as the Schwarzschild black hole and is therefore outlined in Appendix \ref{exactT}. 

\begin{figure}[ht]
\centering 
\includegraphics[width=3.4in]{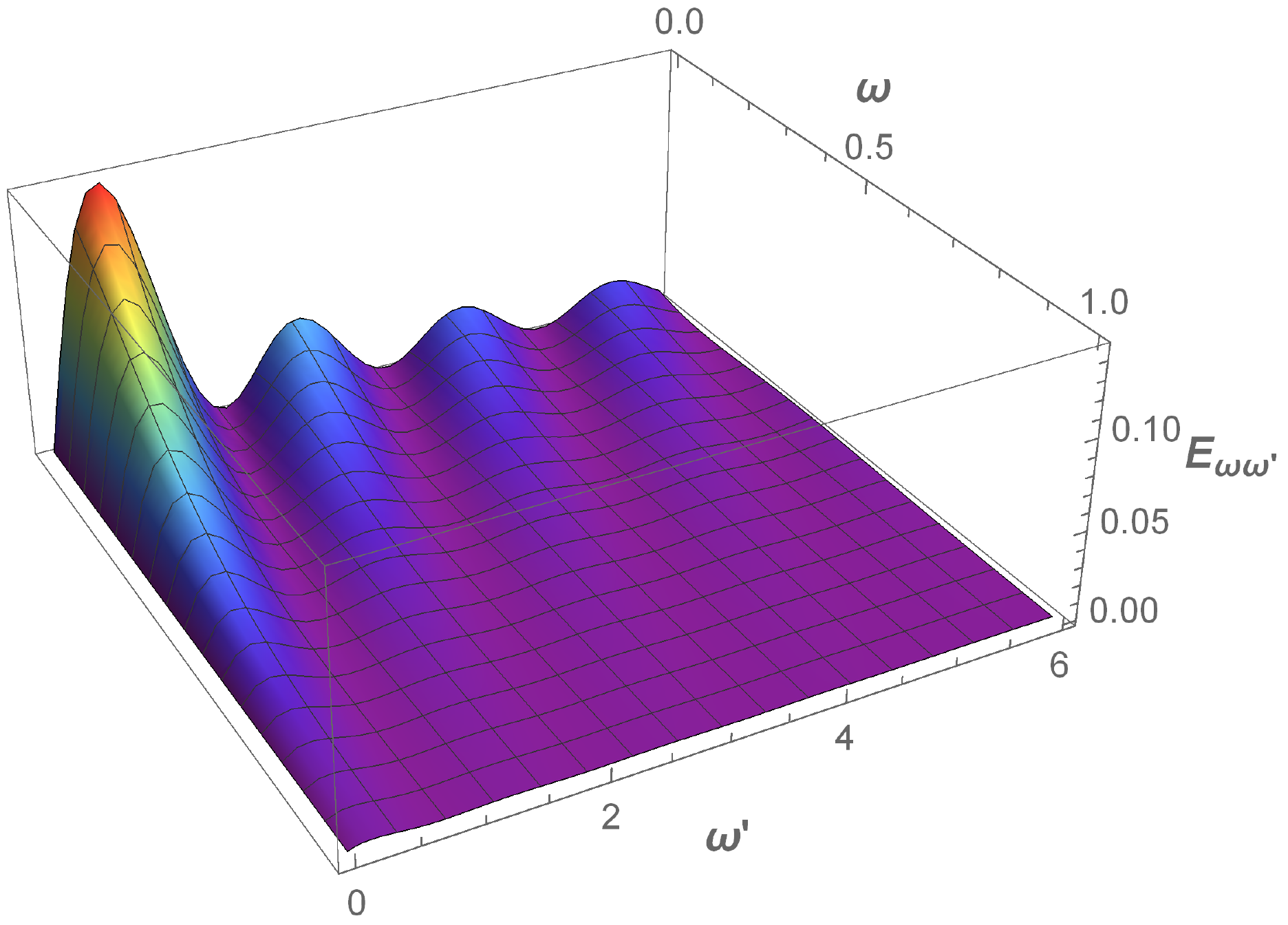} 
\caption{The total energy carried by the particles, Eq.~(\ref{totalenergy}), emitted by the de Sitter mirror is finite.  The integrand $E_{\omega\omega'}$ exhibits dips in energy as seen in $N_{\omega\omega'}$ from Fig.~\ref{fig3}.  There is no infrared divergence, as the limit of $E_{\omega\omega'}$ when $\omega\rightarrow 0$ is $\sin^2(2\omega')/(\pi^2\omega')$.  
}
\label{fig:Energy} 
\end{figure}

Surprisingly, despite infinite acceleration and constant energy flux, Eq.~(\ref{constantflux}), for all times $u$, the two horizons in $v$ appear to conspire to render the total emitted energy, 
\be E = \int_0^\infty d\omega\, \omega N_\omega = \textrm{finite},\label{totalenergy}\ee
finite. The closed form result for $E$ is challenging analytically, but straightforward numerically.  Computing Eq.~(\ref{totalenergy})  for $\kappa =1$ gives $E \approx 5$.  Eq.~(\ref{totalenergy}) is the energy carried by the particles and contrasts with the energy of radiation,
\be E = \int_{-\infty}^\infty du\, F(u) = \textrm{infinite},\ee 
resulting from the quantum stress tensor measured at $\mathscr{I}^+_R$.  A similar finite energy to Eq.~(\ref{totalenergy}) is obtained using a finite-lifetime mirror in \cite{foo2020generating}.  Following Eq.~(\ref{dSthermal}), we plot $E_{\omega\omega'} = \omega N_{\omega\omega'}$ in Fig.~\ref{fig:Energy}.


\section{From de Sitter to Anti-de Sitter}\label{sec:AdS} 

Reference~\cite{Good_2018} showed that constant thermal flux had a 
simple condition when written in terms of proper time (see Appendix~\ref{propertime}) and identified three forms for the 
moving mirror acceleration satisfying this. The first solution is the 
Carlitz-Willey mirror and the second is the de Sitter mirror, the 
focus of this article. The third solution gives the mirror corresponding 
to Anti-de Sitter (AdS) spacetime. This 
eternally gives off negative energy flux, $F = -\kappa^2/(48\pi)$. 

For AdS $f_L=1+r^2/L^2$, which can be thought of as $L^2=1/\Lambda$ being negative. AdS space does not have a horizon, but does have an edge at the origin $r=0$ with the same regularity requirement as the de Sitter ABC.  Taking the de Sitter $\kappa$ imaginary turns $\tanh$ into $\tan$, so  
\be 
f(v)=\frac{2}{\kappa^*}\,\tan^{-1}\frac{\kappa^* v}{2}\ . 
\ee 
This trajectory is the ABC of the AdS spacetime. 
The non-zero beta coefficients are given, by symmetry, through  $\beta_{\omega\omega'}^{dS} = -\beta_{\omega'\omega}^{AdS}$. 
While many mirrors incite episodes of negative energy flux 
(and indeed unitarity requires it for asymptotically static mirrors \cite{Good:2019tnf}), here the total energy is negative. Since the  relationships between particles and energy in quantum field theory are far from resolved, this trajectory provides a good illustration of the pressing issues raised in association with negative energy radiation carried by non-trivial particle production processes.

\section{Conclusions}\label{sec:conc}

We have solved for the particle spectrum produced in de Sitter cosmology by use of the de Sitter moving mirror. This is the 
second of the eternal thermal mirrors (with the first being the Carlitz-Willey solution for Schwarzschild spacetime, and the third, also presented here, corresponding to Anti-de Sitter cosmology). 

The beta Bogoliubov coefficients can be written in 
terms of special functions, and give rise to a particle spectrum  
with a Planck distribution with temperature inversely proportional  to the horizon scale (or proportional to the square root of the cosmological constant). 

The solution has some interesting properties: isolated zeros in the particle count per mode squared despite a thermal particle count spectrum, which  may  be related to destructive 
interference between the double horizons, and finite total energy emission derived via quantum summing, and numerically verified, which contrasts with the infinite energy via an eternally thermal quantum stress tensor. 

The  accelerated boundary correspondence is demonstrated to be a useful tool, enabling us to use the derived de Sitter moving mirror solution to confirm that the distribution of particles produced 
from a de Sitter spacetime is the thermal Planck spectrum, with temperature related to the horizon scale, or alternately mirror acceleration.

\acknowledgments 

MG thanks Joshua Foo and Daiqin Su for stimulating and helpful correspondence.  Funding from state-targeted program ``Center of Excellence for Fundamental and Applied Physics" (BR05236454) by the Ministry of Education and Science of the Republic of Kazakhstan is acknowledged. MG is also funded by the ORAU FY2018-SGP-1-STMM Faculty Development Competitive Research Grant No. 090118FD5350 at Nazarbayev University. 
EL is supported in part by the Energetic Cosmos Laboratory and by 
the U.S.\ Department of Energy, Office of Science, Office of High Energy Physics, under Award DE-SC-0007867 and contract no.\ DE-AC02-05CH11231.

\appendix

\section{Euclidean Method}\label{appx}
Gibbons-Hawking \cite{GibbonsHawking} showed that thermal radiation emanates from the de Sitter horizon, similar to the radiation emanating from the Schwarzschild black hole horizon \cite{Hawking1975} and to the radiation seen by an accelerated observer in the Unruh effect \cite{PhysRevD.14.870}.  Dimensional analysis of the system immediately gives $T \sim 1/L$, while the proportionality factor of $2\pi$ can be obtained via Wick rotation. The static patch has a Euclidean continuation by taking $t_E = i t$, resulting in
\be ds^2 = +f_L dt_E^2 + f_L^{-1} dr^2 + r^2 d\Omega\,.\ee 
The periodicity of Euclidean time, with period $\beta = 2\pi L$ implies a temperature $T = (2\pi L)^{-1}$.  
Essentially, de Sitter space can be viewed as a finite cavity surrounding the observer, with the horizon as its boundary 
\cite{Balasubramanian_2001}.

\section{Proper time dynamics} \label{propertime}

The second of the constant thermal flux solutions in \cite{Good_2018} 
had the proper acceleration written in proper time as 
\be \alpha(\tau) = -\frac{ \kappa}{2}  \tan \frac{\kappa  \tau}{2}\,.\ee 
Converting from $\tau$ to $v$ through $dv/d\tau=d(t+x)/d\tau=\cosh\eta+\sinh\eta$ and 
$\alpha = d\eta/d\tau$ so $\eta=\ln\cos(\kappa\tau/2)$, we find $v=(2/\kappa)\sin(\kappa\tau/2)$ 
so 
\be 
\alpha(v)=-\frac{1}{2}\frac{\kappa^2v}{\sqrt{4-\kappa^2 v^2}}\,, 
\ee 
precisely Eq.~(\ref{a(v)}). Thus the second eternal thermal flux 
solution is indeed equivalent to the de Sitter case. 
An advantage of proper time is that the derivation of constant 
energy flux is particularly simple: 
\be F(\tau) = -\frac{1}{12\pi} \eta''(\tau) e^{2\eta(\tau)} = \frac{\kappa^2}{48\pi}\,,\ee
agreeing with Eq.~(\ref{constantflux}).

\section{Explicit Calculation of Planck Spectra for de Sitter's Moving Mirror}\label{exactT}
 
Thermal behavior is seen by the expectation value:
\be \mathcal{N}_\omega \equiv \int_0^{+\infty} d \omega' \beta^{}_{\omega \omega'}\beta^{*}_{\omega_2 \omega'} = \frac{\delta(\omega-\omega_2)}{e^{2\pi \omega/\kappa}-1}\,.\ee
We derive this starting with the beta coefficients.  After an integration by parts, 
we have
\be \beta_{\omega \omega'} = \frac{2\omega'}{4\pi\sqrt{\omega\omega'}}  \int_{-2/\kappa}^{+2/\kappa}dv_1\, V_1^{-i\omega/\kappa}e^{-i \omega' v_1}, \ee
and its complex conjugate counterpart,
\be \beta^{*}_{\omega_2 \omega'} = \frac{2\omega'}{4\pi\sqrt{\omega_2\omega'}} \int_{-2/\kappa}^{+2/\kappa}dv_2\, V_2^{i\omega_2/\kappa}e^{i \omega' v_2}, \ee
where $V_i \equiv (1+\kappa v_i/2)/(1-\kappa v_i/2)$. 
The spectrum $\mathcal{N}_\omega$ scales as
\be \mathcal{N}_\omega \sim \int d\omega' \int dv_1\int dv_2 V_1^{-i\omega/\kappa}V_2^{i\omega_2/\kappa}  \omega' e^{-i \omega' (v_1-v_2)},\ee
where the proportionality factor is 
$1/(4\pi^2 \sqrt{\omega\omega_2})$. The $\omega'$ integration is done via the introduction of a real $\epsilon >0$ regulator,
\be \int_0^\infty d\omega'\, \omega' e^{-i \omega' Z} = -\frac{1}{(Z- i \epsilon)^2}\,,\ee
giving 
\be \mathcal{N}_\omega =\frac{-1}{4\pi^2\sqrt{\omega\omega_2}} \int dv_1\int dv_2\,   \frac{V_1^{-i\omega/\kappa}V_2^{i\omega_2/\kappa}}{(v_1-v_2-i \epsilon)^2}\,.\ee
A substitution via variables $v_i=(2/\kappa)\tanh 2 S_i$ results in 
\be  \mathcal{N}_\omega = \frac{-1}{4\pi^2\sqrt{\omega\omega_2}}\int dS_1 \int dS_2\, \frac{4 e^{-4 i ( \omega S_1-\omega_2 S_2)/\kappa}}{\sinh^2(2(S_1-S_2))}\,,\ee 
to leading order in small $\epsilon$.  A second substitution simplifies via $Q_{p,m}\equiv S_1\pm S_2$ to 
\be \mathcal{N}_\omega = \frac{-1}{2\pi^2\sqrt{\omega\omega_2}}\int dQ_p e^{-2i \omega_m Q_p/\kappa}\int dQ_m \frac{e^{-2i\omega_p Q_m/\kappa}}{\sinh^2(2Q_m-i\epsilon)}\,,\ee
where $\omega_{p,m} \equiv \omega \pm \omega_2$, and a new  $\epsilon$ has been introduced, and a Jacobian of $1/2$. The first integral is the Dirac delta, so 
\be\mathcal{N}_\omega = \frac{2\pi(\kappa/2) \delta(\omega_m)}{2\pi^2\sqrt{\omega\omega_2}} \int dQ_m \frac{-e^{-2i\omega_p Q_m/\kappa}}{\sinh^2(2Q_m-i\epsilon)}\,.\ee
We can drop the subscript now, $Q_m :=Q$.  For the next integral, we will use the identity 
\be \frac{1}{\sinh^2(\pi x)} =  \sum_{k=-\infty}^{+\infty}\frac{1}{(\pi x+i \pi k)^2}\,,\label{csch}\ee
and now $\omega_p = 2\omega$, to write 
\be \frac{-e^{-2i\omega_p Q/\kappa}}{\sinh^2(2Q-i\epsilon)} = \sum_{k=-\infty}^{+\infty}\frac{-e^{-4 i \omega Q/\kappa}}{(2Q-i\epsilon+  i \pi k)^2}\,. \ee
First integrating, 
\be \sum_{k=-\infty}^{+\infty} \int_{-\infty}^{+\infty} dQ \frac{-e^{-4 i \omega Q/\kappa}}{(2Q-i\epsilon+ i \pi k)^2} = \frac{2\pi \omega}{\kappa} \sum_{k=1}^{+\infty}e^{- 2\pi k\omega/\kappa }\,,\ee 
so that  
\be \mathcal{N}_\omega = \delta(\omega_m)\, \sum_{k=1}^{+\infty} e^{-2\pi k\omega/\kappa}\,,\ee 
we then sum, giving the final result 
\be\mathcal{N}_\omega = \frac{\delta(\omega-\omega_2)}{e^{2\pi \omega/\kappa}-1}\,.\label{Planck}\ee

We can compare to the Schwarzschild mirror \cite{Good_2016}, which has 
beta coefficient squared  
\be N^{S}_{\omega\omega'} :=|\beta_{\omega\omega'}^{\textrm{S}}|^2= \frac{\omega '}{2 \pi \kappa \left(e^{2\pi \omega/\kappa }-1\right) \left(\omega '+\omega \right)^2}\,,\label{Schw}\ee 
with $\kappa=1/(4M)$.  In the high frequency regime, where the modes are extremely red-shifted, $\omega'\gg \omega$, one has the per mode squared spectrum $N_{\omega \omega'} :=|\beta_{\omega\omega'}|^2$ (not $N_\omega$) as 
\be N^{S}_{\omega \omega'} = \frac{1 }{2\pi\kappa \omega'}\frac{1}{e^{\omega/T_s }-1}\,,\label{CWthermal}\ee 
where the Schwarzschild temperature is $T_s=\kappa/(2\pi)$. The de Sitter result is eternally thermal, while the collapse to Schwarzschild black hole is only late-time thermal. The Schwarzschild result for $\mathcal{N}^{S}_{\omega}$ proceeds \cite{Fabbri} to the penultimate result 
\be \mathcal{N}^{S}_\omega = \frac{\kappa}{2\pi \omega} \left|\Gamma\left(1+i\frac{\omega}{\kappa}\right)\right|^2 e^{-\pi \omega/\kappa} \delta(\omega_m)\,,\ee
which reduces to $\mathcal{N}^{S}_\omega = \delta(\omega_m)(e^{2\pi \omega/\kappa}-1)^{-1}$, identical in form to Eq.~(\ref{Planck}).

\bibliography{main}

\end{document}